\documentstyle[aps,eqsecnum]{revtex}
\begin{document}
\newtheorem{algorithm}{PDP Algorithm}
\newtheorem{definition}{Definition}
\newcommand{\dia}{\mbox{diag} }
\newcommand{\tr}{\mbox{Tr}\ }

\newcommand{\aaa}{A_\alpha}

\newcommand{\ha}{H_\alpha}
\newcommand{\gab}{g_{\alpha\beta}}
\newcommand{\gba}{g_{\beta\alpha}}
\newcommand{\la}{\Lambda_\alpha}
\newcommand{\va}{V_\alpha}
\newcommand{\be}{\begin{equation}}
\newcommand{\ee}{\end{equation}}
\newcommand{\calh}{{\cal H}}
\newcommand{\ra}{\rho_\alpha}
\newcommand{\ca}{{\cal A}}
\draft
\twocolumn[
\hsize\textwidth\columnwidth\hsize\csname@twocolumnfalse\endcsname\title
{Event-Enhanced Quantum Theory\\ {\large and} \\
Piecewise Deterministic Dynamics}
\author{Ph.~Blanchard${}^\flat$ \ and\ A.~Jadczyk${}^\sharp$}
\address{${}^\flat$ Faculty of Physics and BiBoS,
University of Bielefeld\\
Universit\"atstr. 25,
D-33615 Bielefeld\\
${}^\sharp$ Institute of Theoretical Physics,
University of Wroc{\l}aw\\
Pl. Maxa Borna 9,
PL-50 204 Wroc{\l}aw}
\maketitle
\begin{abstract}
The standard formalism of quantum theory is enhanced and
definite meaning is given to the concepts of experiment, measurement
and event. Within this approach one obtains a
uniquely defined piecewise deterministic algorithm generating quantum
jumps, classical events and histories of single quantum objects. The
wave-function Monte Carlo method of Quantum Optics is generalized and
promoted to the level of a fundamental process generating all the real
events in Nature. The already worked out applications include SQUID-tank
model and generalized cloud chamber model with GRW spontaneous localization
as a particular case. Differences between the present approach and quantum
measurement theories based on environment induced master equations are
stressed.
Questions: what is classical, what is time, and what are observers are
addressed. Possible applications of the new approach are suggested,
among them connection between the stochastic commutative geometry and Connes'
noncommutative formulation of the Standard Model, as well as potential
applications to the theory and practice of quantum computers.
\end{abstract}
]
\section{Introduction}
Quantum Mechanics occupies a very particular place among scientific
theories; indeed it is at once one of the most successful and one
of the most mysterious ones. Its success lies
undoubtedly in the fact that using Quantum Mechanics one can
predict properties of atoms, of molecules, of chemical reactions,
of conductors and insulators and much more. These
predictions were confirmed by precise measurements and by
the technological  progress that is based on quantum
phenomena.
The mystery resides in the problem of interpretation of Quantum
Theory - which does not follow from the formalism itself but is left to
discretion of a physicist. As a result, there is still no general agreement
about how Quantum Mechanics is best understood and to what extent
it can be considered as exact and complete.\\
As emphasized already by E. Schr\"odinger \cite{schrod55} what is
definitively and completely missing in Standard Quantum Mechanics is
an explanation of experimental facts, as it does not tell us how to
generate time series of events recorded during real experiments
on {\em single individual systems}.
H.P. Stapp \cite{stapp75,stapp77,stapp85a,stapp93}
and R. Haag \cite{haag90,haag92} emphasized the role and importance
of \lq {\em events}\rq\, in quantum physics. J. Bell \cite{bell87a}
stressed the fundamental necessity of distinguishing
\lq {\em definite events}\rq\, from \lq {\em just wavy possibilities}\rq .

In 1969 E.B. Davies
\cite{ebdavies69} (see also \cite{ebdavies76})
introduced the \lq {\em space of events}\rq\/ in his
mathematical theory of quantum stochastic processes which extended
the standard formalism of quantum theory. His theory went beyond a standard
quantum
measurement theory and, in its most general form,  was not expressible in
terms of quantum master equations alone. Later on Srinivas, in a joint
paper with  Davies \cite{srinivas81},  specialized Davies' general and
mathematically sophisticated scheme
to photodetection processes. Photon counting
statistics predicted by this theory were successfully verified
in fluorescence experiments which caused R. J. Cook to revisit
the question \lq {\em what are quantum jumps}\rq\/ \cite{cook88}.
A related question: \lq{\em are there quantum jumps}\rq\/ was asked
by  J. Bell \cite{bell87b} in connection with the
idea of spontaneous localization put forward by Ghirardi, Rimini
and Weber \cite{grw86}.

In the eighties quantum optics experiments started to call for
efficient methods of solving quantum master equations that
described effective coupling of atoms to the radiation modes.
The works of Carmichael \cite{carm93},  Dalibard, Castin and M{\o}lmer
\cite{dal92,molmer93}, Dum, Zoller and Ritsch \cite{dum92},
Gardiner, Parkins and Zoller \cite{gard92},
developed Quantum Monte Carlo (QMC) algorithm for simulating solutions
of master equations.\footnote{Less general scheme was proposed by Teich and
Mahler \cite{teich92} who
tried to extract a specific jump process directly from the orthogonal
decomposition of time evolving density matrix.}
 The algorithm emerged from the the seminal papers  of
Davies \cite{ebdavies69,ebdavies76} on quantum stochastic processes, that
were followed
by numerous works on photon counting and continuous measurements
(cf. Refs. \cite{srinivas84,barch85,barch91}).
 It was soon realized
(cf. e.g. \cite{gisin84,gisin89,diosi88,diosi89,pearle89})
that the same master
equations can be simulated either by Quantum Monte Carlo method
based on quantum jumps, or by a continuous quantum state diffusion.
Wiseman and Milburn \cite{wise93}
discussed the question of which experimental detection schemes
are better described by continuous diffusions rather than by
discontinuous jump simulations. The two approaches were recently
put into comparison also by Garraway and Knight \cite{garr94},
while Gisin et al. \cite{gisin93} argued that \lq {\em the quantum jumps
can be clearly seen}\rq\/ also in the quantum state diffusion plots.
Apart from the numerical usefulness of quantum jumps and empirical
observability of photon counts, the debate of their \lq{\em
reality}\rq\/ continued. A brief synthesis of the present state of
the debate has been given by M{\o}ller in the final paragraphs of
his 1994 Trieste lectures \cite{moller94}:
\begin{quotation}
The macroscopic collapse has been explained, the elementary collapse,
however remains as an essential and unexplained ingredient of the
theory.\\
A real advantage of the QMC method: We can be sitting there and
discussing its philosophical implications and the deep questions of
quantum physics while the computer is cranking out numbers which we
need for practical purposes and which we could never obtain in any
other way. What more can we ask for?
\end{quotation}

In the present paper we argue that indeed \lq {\em more}\rq\, can be
not only asked for, but that it can be also provided.
The picture that we propose developed from
a series of papers
\cite
{blaja93a,blaja93b,blaja93c,blaja93d,blaja94b,ja94a,ja94b,ja94c,blaja94c},
where we treated
several applications including SQUID--tank \cite{blaja93c} and
cloud chamber model (with GRW spontaneous localization)
\cite{ja94b,ja94c,blaja94c}.
In the sequel we will refer
to it as Event Enhanced Quantum Theory (EEQT). EEQT is a minimal extension
of the standard quantum theory that accounts for events.
In the next three sections
we will describe formal aspects of EEQT, but we will attempt to reduce the
mathematical apparatus to the absolute minimum. In the final Sect. 4
we will propose to use EEQT for describing not only quantum measurement
experiments, but all the real processes and events
in Nature.
The new formalism rises new questions, and in Sect. 4 we will point out
some of these questions. One of the problems that can be discussed in a
somewhat new light is that of the role of \lq observers\rq\/ and IGUS-es
(using terminology of Gell--Mann and Hartle, cf. \cite{gell89}). We will
also make a comment on a possible interpretation of  Connes' version of
the Standard Model as a stochastic geometry a'la EEQT, with jumps
between the two copies of space--time. We will also mention relevance
of EEQT to the theory and practice of quantum computers. The reader
interested in the results and perspectives rather than in the mathematical
formulation may skip Sections 2 and 3.
\subsection{Summary}
Using informal language EEQT can be summarized as follows:
Given a \lq wavy\rq\ quantum system $Q$ we allow it to generate distinct
classical traces - {\em events}. Quantum wave functions are not directly
observable. They may be considered as hidden variables of the theory.
Events are discrete and real. Typically one can think of detection events
and pointer readings in quantum mechanics, but also of creation--annihilation
events in quantum field theory. They can be observed but they do not need an
observer for their generation (although {\em some} may be triggered by
observer's participation). They are either recorded or they
are causes for other events. It is convenient to represent events
as changes of state of a suitable  classical system. Thus formally
we divide the world into $Q\times C$ -- the quantum and the classical
part. They are coupled together via a specific dynamics that
can be encoded in a Liouville evolution equation for statistical states
of the total $Q\times C$ system. To avoid misunderstanding we wish
to stress it rather strongly: the fact that $Q$ and $C$ are
coupled by a dissipative rather than unitary dynamics does not mean that
noise, or heat, or chaos, or environment, or lack of knowledge, are involved.
In fact each of these factors, if present -- and all of them are
present in real circumstances, only blurs out
transmission of information between $Q$ and $C$. The fact that $Q$ and
$C$ must be coupled by a dissipative rather than by reversible
dynamics follows from no--go theorems that are based on rather
general assumptions \cite{land91,ozawa92,ja94a}. We go beyond these
abstract no--go theorems that are telling us what is not possible.
We look for what is possible, and we propose a class of couplings that,
as we believe, is optimal for the purposes of control and measurement.
With our class of couplings no more dissipation is introduced than it is
necessary for transmission of information from $Q$ to $C$. Thus our
Liouville equation that encodes the measurement process is to be
considered as {\em exact}\/, not an approximate one (adding noise to
it will make it approximate). Given such
a coupling we show that the Liouville equation encodes in a unique
way the algorithm for generating admissible histories of individual
systems. This algorithm generalizes the one of Davies \cite{ebdavies69}
as well as descending from the Davies' theory wave-function Monte Carlo
method.\footnote{The wave-function Monte Carlo method of Quantum Optics
may be considered as
mutilated version of the general Davies' process. Indeed, Davies'
\lq events\rq\/ are
forgotten there, and only the Liouville equation is accepted. That leads to
arbitrariness of choosing between different jump or diffusion processes,
arbitrariness which was not present in Davies' original theory.}
The algorithm describes joint evolution of an individual $Q\times C$ system
as a piecewise deterministic process. Periods of continuous deterministic
evolution are interrupted by die tossings and random jumps that are
accompanied by changes of state of $C$ - events.
 We call it Piecewise
Deterministic Process Algorithm, in short PDP
(the term PDP has been introduced by M.H.A. Davis \cite{davis93}).
The algorithm is probabilistic what reflects the fact that the
quantum world although governed by deterministic Schr\"odinger equation
is, as we know from experience, {\em open}\/ towards the classical world of
events, and the total
system $Q\times C$ is thus open towards the future. The PDP algorithm
identifies the probabilistic laws according to which
times of jumps and the events themselves are chosen.
Our generalized framework
enables us not only to gain information about the quantum system
but also to utilize it by a feed--back control of the $Q\times C$ coupling
i.e. making the coupling dependent on the actual state of the classical
system (which may depend on a record of previous events).

Briefly, our Event--Enhanced formalism can be described as follows:
to define an {\em experiment} we must start with a division $Q\times C$.
Assuming, for simplicity, that $C$ has only finite number of states
(which may be thought of as \lq pointer positions\rq\,)
$\alpha=1,\ldots ,m$, we define {\em event} as a {\em change of state}\/
 of $C$.
Thus there are $m^2-m$ possible events. An experiment is then described
by a completely positive coupling $V$ of $Q$ and $C$. A
coupling is specified by $(i)$ a family $H$ of quantum Hamiltonians
$H_\alpha$
parametrized by the states of $C$, $(ii)$ a family $V$ of $m^2-m$ of quantum
operators $\gab$, with $g_{\alpha\alpha}\equiv 0$. In Refs.
\cite
{blaja93a,blaja93b,blaja93c,blaja93d,blaja94b,ja94a,ja94b,ja94c,blaja94c}
we have described simple general rules for constructing $\gab$-s, and we
described non-trivial examples, including SQUID-tank model and
generalized \lq cloud chamber\rq\/ model that covers GRW spontaneous
localization model
as a particular, homogeneous, case.
The self--adjoint operators $H_\alpha$ determine the unitary
part of quantum evolution between jumps, while $\gab$
determine jumps, their rates and their probabilities, as well as the
non-unitary
contribution to the continuous evolution between jumps. As an example, in
the SQUID--tank model \cite{spiller92} the variable $\alpha$ is the
flux through the
coil of the classical radio--frequency oscillator circuit, and it affects,
through a transformer, the SQUID Hamiltonian. $\gab$ have there also
very simple meaning \cite{blaja93c}, as the shifts of the classical circuit
momentum caused by a (smoothed out, operator--valued) quantum flux.

Time evolution of
statistical states of the total $Q\times C$ system is described by the
Liouville equation:
\be
{\dot \rho}_\alpha=-i\left[\ha,\ra\right]+
\sum_\beta \gab\, \rho_\beta\, \gab^\star
-{1\over 2}\{ \la,\ra\},
\ee
where
\be
\la=\sum_\beta \gba^\star\, \gba.
\ee
The operators $\ha$ and $\gab$ can be allowed to
depend explicitly on time, so that the coupling can be switched
on and off in a controlled way. Moreover, to allow for phase transitions
the quantum Hilbert space may  change with $\alpha$.
We show in Sect. 2 that the above Liouville equation determines
a piecewise deterministic process that generates histories
of individual systems. In Sect. 3 we provide argument showing that within
our framework the process is unique. Our PDP is given by the following simple
algorithm which generalizes that of QMC:\footnote{Notice that in the
$\ha$ and $\gab$ in the algorithm may explicitly depend on time.}
\begin{algorithm}
Let us assume a fixed, sufficiently small, time step $dt$.
Suppose that at time $t$ the
system is described by a quantum state vector $\psi$ and a classical
state $\alpha$. Compute the scalar product
$\lambda(\psi,\alpha)=<\psi,\la\,\psi>$.
Then choose a uniform random number $p\in [0,1]$, and jump
if $p<\lambda(\psi,\alpha)dt$.
When jumping, change $\alpha\rightarrow
\beta$ with probability $p_{\alpha\rightarrow\beta}=
\Vert\gba\psi\Vert^2/\lambda(\psi,\alpha)$, and change
$\psi\rightarrow\gba\psi/
\Vert\gba\psi\Vert$. If not jumping, change

$$
\psi \rightarrow
{{\exp \{-i\ha dt-{1\over2}\la dt\}\psi}\over{\Vert
\exp \{-i\ha dt-{1\over2}\la dt\}\psi\Vert}} ,\quad
t \rightarrow  t+dt.
$$ Repeat the steps.\footnote{There are
several methods available for
efficient computation of the exponential for $dt$ small enough
-- cf. Ref. \cite{raedt87}.}
\end{algorithm}
EEQT proposes that the PDP Algorithm describes {\em in an exact way}
all real events as they occur in Nature, provided
we specify correctly $Q,C,H$ and $V$.
More on this subject can be found in Sect. 4.
In the following section we will formulate more precisely the basic structure
of EEQT.

\section{Formal scheme of  EEQT}

Let us briefly describe the mathematical framework that we use. To define
events, we introduce a classical system $C$, and possible events will be
identified with changes of (pure) state of $C$.  To concentrate on main
ideas rather than on technical details we will  consider the simplest
situation corresponding to a finite set of possible events. It is possible
and necessary in many applications to handle infinite dimensional
generalizations of this framework. The space of states ${\cal S}_c$ has $m$
states, denoted by $\alpha=1,\ldots,m$. These are the pure states of $C$.
Statistical states of $C$ are probability measures on ${\cal S}_c$ - in
our case just sequences $p_\alpha\geq 0, \sum_\alpha p_\alpha=1$. We
will also consider the algebra of (complex) observables of $C$. This will be
the algebra $\ca_c$ of complex functions on ${\cal S}_c$ - in our case
just sequences $f_\alpha, \alpha =1,\ldots,m$ of complex numbers. It is
convenient to use Hilbert space language even for the description of that
simple classical system. Thus we introduce $m$-dimensional Hilbert space
$\calh_c$ with a fixed basis, and realize $\ca_c$ as the algebra of
diagonal matrices $F=\dia(f_1,\ldots,f_m)$. Statistical states of $C$ are
then diagonal density matrices $\dia(p_1,\ldots,p_m)$, and pure states of
$C$ are vectors of the fixed basis of $\calh_c$. Events are ordered
pairs of pure states $\alpha\rightarrow\beta$, $\alpha\neq\beta$. Each
event can thus be represented by an $m\times m$ matrix with $1$ at the
$(\alpha,\beta)$ entry, zero otherwise. There are $m^2-m$ possible events.
Statistical states are concerned with ensembles, while pure states and events
concern individual systems.

Let $Q$ be the quantum system whose observables are from the algebra
$\ca_q$ of bounded operators on a Hilbert space $\calh_q$.
Its pure states are unit vectors in $\calh_q$ with understanding
that proportional vectors describe the same quantum state. Its statistical
states are given by non--negative density matrices ${\hat\rho}$,  with
$\tr ({\hat\rho})=1$. Then pure states can be identified with those density
matrices that are idempotent ${\hat\rho}^2={\hat\rho}$, i.e. with
one-dimensional orthogonal projections.

Let us now consider the total system $T=Q\times C$. First, we consider its
statistical description, only after that we will discuss the dynamics and the
coupling of $Q$ and $C$. For the algebra $\ca_t$ of observables of $T$
we take the tensor product of algebras of $C$ and $Q$:
$\ca_t=\ca_q\otimes\ca_c$. Thus
$\ca_t$ can be thought of as algebra of {\em diagonal} $m\times m$
matrices $A=(a_{\alpha\beta})$, whose entries are quantum operators:
$a_{\alpha\alpha}\in \ca_q$, $a_{\alpha\beta}=0$ for $\alpha\neq\beta$.
The classical and quantum algebras are then subalgebras of $\ca_t$;
$\ca_c$ is realized by putting $a_{\alpha\alpha}=f_\alpha I$, while
$\ca_q$ is realized by choosing $a_{\alpha\beta}=a\delta_{\alpha\beta}$.
Statistical states of $Q\times C$ are given by $m\times m$ diagonal matrices
$\rho=\dia(\rho_1,\ldots,\rho_m)$ whose entries are positive operators
on $\calh_q$, with the normalization $\tr (\rho)=\sum_\alpha\tr
(\ra)=1$. Tracing over $C$ or $Q$ produces the effective states of $Q$ and
$C$ respectively: ${\hat\rho}=\sum_\alpha \ra$, $p_\alpha=\tr (\ra )$.
Duality between observables and states is provided
by the expectation value $<A>_\rho=\sum_\alpha \tr (\aaa\ra)$.

We consider now dynamics. Quantum dynamics, when no information is
transferred from $Q$ to $C$, is described by Hamiltonians $\ha$,
that may depend on the actual state of $C$ (as indicated by the index
$\alpha$). We will use matrix notation and write $H=\dia(\ha)$.
A {\em coupling} of $Q$ to $C$ is specified by a matrix
$V=(\gab)$, with $g_{\alpha\alpha}=0$. To transfer  information
from $Q$ to $C$ we need a non--Hamiltonian term which provides
a completely positive (CP) coupling. We propose to consider
couplings for which the evolution
equation for observables and for states is given by the Lindblad form:
\be
{\dot A}=i[H,A]+{\cal E}\left(V^\star AV\right)-{1\over2}\{\Lambda,A\},
\ee
\be
{\dot \rho}=-i[H,\rho]+{\cal E}(V\rho V^\star)-{1\over2}\{\Lambda,\rho\},
\ee
where
${\cal E}:(A_{\alpha\beta})\mapsto \dia (A_{\alpha\alpha})$ is the
conditional expectation
onto the diagonal subalgebra given by the diagonal projection, and
\be
\Lambda={\cal E}\left(V^\star V\right).
\ee
We can also write it down in a form not involving ${\cal E}$:
\be
{\dot A}=i[H,A]+\sum_{\alpha\neq\beta}V_{[\beta\alpha]}^\star
AV_{[\beta\alpha]}-{1\over2}\{\Lambda,A\},
\ee
with $\Lambda$ given by
\be
\Lambda=\sum_{\alpha\neq\beta}V_{[\beta\alpha]}^\star V_{[\beta\alpha]},
\ee
and where $V_{[\alpha\beta]}$ denotes the matrix that has only one
non--zero entry, namely $\gab$ at the $\alpha$ row and $\beta$ column.
Expanding the matrix form we have:
\be
{\dot A}_\alpha=i[\ha,\aaa]+\sum_\beta \gba^\star
A_\beta \gba - {1\over2}\{\la,\aaa\},\label{eq:lioua}
\ee
\be
{\dot \rho}_\alpha=-i[\ha,\ra]+\sum_\beta \gab
\rho_\beta \gab^\star - {1\over2}\{\la,\ra\},\label{eq:liour}
\ee
where
\be
\la=\sum_\beta \gba^\star \gba.
\ee
Again, the operators $\gab$ can be allowed to depend explicitly
on time.\footnote{When $V=V^\star$ i.e. when $\gab=\gba^\star$, then our
coupling satisfies the so called {\em detailed balance
condition} - cf. \cite{maj84,maj94}.}

Following \cite{ja94c} we now define {\em experiment} and
{\em measurement}:

\begin{definition}
An {\bf experiment} is a CP coupling between a quantum and a classical
system. One observes then the classical system and attempts to learn
from it about characteristics of state and of dynamics of the quantum
system.
\end{definition}

\begin{definition}
A {\bf measurement} is an experiment that is used for a particular
purpose: for determining values, or statistical distribution of values,
of given physical quantities.
\end{definition}

\noindent{\bf Remark}\,
The definition of experiment above is concerned with the {\em conditions}
that define it. In the next sections we will derive the PDP
algorithm that simulates a typical {\em run} of a given experiment. In
practical situations it is rather easy to decide what constitutes $Q$,
what constitutes $C$ and how to write down the coupling. Then, if
necessary, $Q$ is enlarged, and $C$ is shifted towards more macroscopic
and/or more classical. However the new point of view that we propose
allows us to consider our whole Universe as \lq experiment\rq\,
and we are witnesses and participants of one particular run. Then the
question arises: {\em what is the true} $C$? We will comment on this question
in the closing section.

\section{From the Liouville equation for ensembles to the PDP
algorithm for single systems}
\subsection{Derivation of the PDP}
Instead of constructing the PDP out of the Liouville equation, we
will show that Eq. (\ref{eq:lioua}) is compatible with
 the PDP Algorithm
described in Section 1. Then, in the next subsection we will give
arguments that can be used for proving its uniqueness. In order to
prove compatibility of the Liouville equation  (\ref{eq:lioua})
with the PDP Algorithm, i.e. to show that  (\ref{eq:lioua}) follows from PDP
by averaging, we
will use the theory of piecewise deterministic processes (PDP)
developed by M.H.A. Davis
\cite{davis93}. By Theorem (26.14) of Ref. \cite{davis93} our PDP
Algorithm leads to the following infinitesimal generator ${\cal D}$
acting on complex valued functions $f(\psi,\alpha)$\footnote{If
$H$ or $V$ explicitly depend on time, then we should add time $t$ as
the third argument of $f$.}
\be
\begin{array}{ll}
{\cal D} f (\psi,\alpha)={\cal Z}f(\psi,\alpha)+&\null\\
+\lambda(\psi,\alpha)
\sum_\beta\int_\phi (f(\phi,\beta)-f(\psi,\alpha)){\cal K}(d\phi,\beta;
\psi,\alpha)&\null ,\label{eq:gen}
\end{array}
\ee
where
\be
{\cal Z}f(\psi,\alpha)={d\over dt}
f\left({\exp(-iH_\alpha-{1\over2}\la)\psi\over\Vert
\exp(-iH_\alpha-{1\over2}\la)\psi\Vert
},\alpha\right)|_{t=0},
\ee
\be
\lambda(\psi,\alpha)=<\psi,\la\psi>,
\ee
and
\be
{\cal K}(d\phi,\beta;\psi,\alpha)=
{\Vert \gba\psi\Vert^2\over{\lambda(\psi,\alpha)}}\ \delta\left(
\phi-{\gba\psi\over\Vert\gba\psi\Vert}\right)\, d\phi .
\ee
The above formula holds for {\em arbitrary}\/ functions of $\psi$
and $\alpha$.
However, because $Q$ is quantum rather than classical, and because we
are interested only in {\em linear} quantum mechanics, we need to consider
only
very special class of functions of $\psi$, namely those given by expectation
values of linear quantum observables.\footnote{As is well known, quantum
mechanics can be considered as a particular case of classical mechanics,
namely as (in general -- infinite--dimensional) classical mechanics with a
restricted set of observables.} To this end for each observable $A$ of
the total system we associate function $f_A(\psi,\alpha)$ defined by its
expectation value: $f_A(\psi,\alpha)=<\psi,\aaa\psi>$. Then, sandwiching
the Liouville  equation (\ref{eq:lioua}) between two $\psi$ vectors,
one can check (essentially by inspection) that its right hand side can be
written up exactly as in Eq. (\ref{eq:gen}) for $f=f_A$. That proves that
our Liouville equation follows from the PDP Algorithm. Examples and
 details of the computation can be found in
Refs \cite{blaja93b,ja94a,ja94b}.
\subsection{Uniqueness of the PDP}
In ordinary, i.e. non--enhanced by events, quantum theory there will
be many random processes on the unit ball of the Hilbert space that
reproduce the same master equation for density matrix. The reason for
this non--uniqueness being the fact that the convex set of statistical
states of a quantum system is, contrary to the classical case, not a
simplex. That is a given density matrix will decompose in infinitely
many ways into pure states. (The fact that in a non--degenerate case
there is a preferred orthogonal decomposition is just a mathematical
artifact that has no statistical justification.) This non--uniqueness
is equivalent to another fact, namely in quantum theory we have at our
disposal not all functions $f(\psi)$ of pure states, but only those
given by expectation values of linear observables $f_A(\psi)=<\psi,A\psi>$.
The Liouville equation gives us time evolution, and thus its infinitesimal
generator only on such functions - special polynomials in $\psi$ of degree 2,
while to reconstruct the random process in $\psi$ space we need to know time
evolution of characteristic functions of sets. Thus we have to extend our
generator from functions $f_A(\psi)$ given by linear observables to arbitrary
functions $f(\psi)$. Such an extension is non--unique and different
extensions give rise to different random processes.

The situation is different when we discuss not arbitrary quantum master
equations but experiments and measurements
in EEQT. Here we have $Q$ {\em and}\/ $C$, and a special form of a Liouville
equation - that given by Eq. (\ref{eq:lioua}). As we already remarked, it
describes transfer of information from $Q$ to $C$ without introducing
unnecessary (and harmful for the data) dissipation - that is why there
should be zeros on the
diagonal of the coupling $V$--matrix. That particular form of the
Liouville equation has, as we will show now, a very special
property. Namely, starting with a pure state $(\psi,\alpha)$ of the
total system, after time $dt$ we have a mixed state; there will be
mixing along classical - which is uniquely decomposable, and there
will be mixing along quantum - which decomposes nonuniquely. However,
while mixing along classical is of the order $dt$, mixing along quantum
is only of the order $dt^2$. That is the special property that allows for a
unique determination of the random process in infinitesimal steps. It is
from this property that one can see again that our dissipation {\em is not
caused by quantum noise}\/ - rather it is the necessary minimal price that
must be paid for any bit of information recived from the quantum
system.\footnote{Quantum noise (cf. Ref. \cite{gard91}), if present,
it would appear on the
diagonal of the $\gab$ matrix, and we have put it explicitly to zero.}

To see the last point explicitly, we  use Eq. (\ref{eq:liour}) to
compute $\ra (dt)$ when the initial state $\ra (0)$ is pure:
\be
\ra(0)=\delta_{\alpha\alpha_0}|\psi_0><\psi_0|.
\ee
 In the equations
below we will discard terms that are higher than linear order in $dt$.
For $\alpha=\alpha_0$ we obtain:
\be
\begin{array}{lrl}
\rho_{\alpha_0}(dt)=&
|\psi_0><\psi_0|&-i[H_{\alpha_0},|\psi_0><\psi_0|]\, dt- \\
\null&\null&-{1\over2}\{\Lambda_{\alpha_0},|\psi_0><\psi_0|\} \, dt\/,
\end{array}
\ee
while for $\alpha=\alpha_0$
\be
\rho_{\alpha_0}(dt)=
g_{\alpha\alpha_0}|\psi_0><\psi_0|g_{\alpha\alpha_0}^\star\, dt
\ee
The term for $\alpha=\alpha_0$ can be written as
\be
\rho_{\alpha_0}(dt)=p_{\alpha_0}|\psi_{\alpha_0}><\psi{\alpha_0}|,
\label{eq:a0}
\ee
where
\be
\psi_{\alpha_0}={
{\exp\left(-iH_{\alpha_0}dt-
{1\over2}\Lambda_{\alpha_0}dt\right)\psi_0}
\over{\Vert\exp\left(-iH_{\alpha_0}dt-
{1\over2}\Lambda_{\alpha_0}dt\right)\psi_0 \Vert}},
\ee
and
\be
p_{\alpha_0}=1-\lambda(\psi_0,\alpha_0 )dt.
\ee
The term with $\alpha\neq\alpha_0$ can be written as:
\be
\ra (dt)=p_\alpha \, |\psi_\alpha><\psi_\alpha|\, ,\label{eq:a}
\ee
where
\be
p_\alpha=\Vert g_{\alpha\alpha_0}\psi_0\Vert^2 dt,
\ee
and
\be
\psi_\alpha={
{g_{\alpha\alpha_0}\psi_0}\over{\Vert g_{\alpha\alpha_0}\psi_0\Vert}}
\ee
This representation is unique and it defines the infinitesimal version of
our PDP.
\section{Where are we now?}
We have seen that Quantum Theory can be enhanced in a rather simple way
to predict new facts and to streighten old mysteries.
The EEQT that we have outlined above has several important
advantages. One such advantage is of practical nature: we may use the
algorithms it provides and ask
computers \lq to crank out numbers that are needed in experiments and that
can not
be obtained in another way\rq\/ .
For example in \cite{blaja93c} we have shown
how to generate pointer readings in a tank circuit coupled to a SQUID,
while in \cite{ja94b,ja94c} the algorithm generating detection events
of an arbitrary geometrical configuration of particle position detectors
was derived. As a particular case, in a continuous homogeneous, limit we
have reproduced GRW spontaneous localization model.
Many other examples come from quantum optics, since QMC
is a special case of our approach, namely when {\em events} are
not feed--backed into the system and thus do not really matter.

Another advantage of EEQT is of a conceptual nature: in EEQT
we need only one postulate: {\em that events can be observed}\,. All
the rest can and should be derived from this postulate. All probabilistic
interpretation, everything that we have learned about eigenvalues,
eigenvectors, transition probabilities etc. can be derived from the
formalism of EEQT. Thus in \cite{blaja93a} we have shown that
probability distribution of eigenvalues of Hermitian observables
can be derived from the simplest coupling, while in \cite{blaja94c}
we have shown that Born's interpretation can be derived from the simplest
possible model of a position detector. Moreover, in \cite{ja94a} it was shown
that EEQT can also give definite predictions for non--standard
measurements, like those involving noncommuting operators.

It is also possible that using the ideas of EEQT may throw a new
light into some applications of non--commutative geometry. Namely, when
$C$ consists of {\em two points}, then our $V$ can be interpreted as
Quillen's
superconnection (cf. \cite{coq91} and references there).
Indeed, with $g_{10}=\phi M, g_{01}={\bar \phi}M^\dagger$,
our $V$ of Section 2 plays the same role as $D_\nabla$ operator in Connes'
noncommutative gauge theory \cite[Section 2]{connes90}.
That suggests that Connes' $Z^2$-graded
non--commutative geometry version of the Standard Model can be interpreted
and understood as a {\em commutative but stochastic geometry},
with continuous parallel transport (determined by gauge fields) interrupted
by random jumps between two copies of the universe (determined by Higgs
fields), as in the PDP algorithm.\footnote{Cf. also \cite{ebdavies94} for
relation between superconnections and classical Markov processes.}

Another potential field of application of EEQT is in the theory and
practice of quantum computation. Computing with arrays of coupled
quantum rather than classical systems seems to offer advantages
for special classes of problems (cf. \cite{lloyd93} and references
therein). Quantum computers will have however to use classical
interfaces, will have to communicate with, and be controlled by
classical computers. Moreover,
we will have to understand what happens during individual runs. Only EEQT
is able to provide an effective framework to handle these problems. It
kepps perfect balance of probabilities without introducing
\lq negative probabilities\rq\/, and it needs only standard random number
generators for its simulations.
For a recent work  where similar ideas are considered cf. \cite{korn93}

EEQT is
a precise and predictive theory. Although it appears to be correct, it
is also yet incomplete.
The enhanced formalism and the enhanced framework not only
give enhanced answers, they also
invite asking new questions. Indeed, we are tempted to consider the
possibility that PDP can be applied not only to what we call experiments,
but also, as a \lq world process\rq\,  to the entire universe
(including all kinds of \lq observers\rq ). Thus we may assume
that all the events that happened were generated by a particular PDP
process, with some unknown $Q$,$C$,$H$ and $V$. Then, assuming that past
events are known, the future is
partly determined and partly open. Knowing $Q,C,H,V$ and knowing
the actual state (even if this knowledge is fuzzy and uncertain), we are
in position to use the PDP algorithm to generate probable
future series of events. With such a promotion of the PDP to the role of
a universal world process
questions arise that could not be asked before: {\em what is $C$ and what
is $V$?}\, , and perhaps also: {\em what is $t$?}\, and
{\em what are \lq we\rq}. Of course we
are not in a position to provide answers. But we can discuss possibilities
and we can provide hints.
\subsection{What is time?}
Let us start with the question: {\em what is time}? Answering that
time is determined by the thermodynamic state of the system \cite{connes94}
is not enough, as we would like to know how did it happen that a particular
thermodynamic state has evolved, and to understand this we must assume
evolution, and thus we are back with the question: what is time if not just
counting steps of this evolution. We are tempted to
answer: {\em time is just a measure of the number of events that happened
in a given place}. If so, then time is discrete, and there is another
time, that counts the deterministic steps between events. In that case
die tossing to decide whether the next step is to be an event or not is
probably uneconomic and unnecessary; it is quite possible that the Poissonian
character of events is a result of some ergodic theorem, when we
use not the \lq true\rq\, discrete time, but some continuous \lq averaged\rq\,
time (averaged over a neighborhood of a given place). Thus a possible
algorithm for a finite universe would be discrete, with die tossing
every $N$ steps, $N$ being a fixed integer, and continuous, averaged time
would appear only in a thermodynamic limit. In fact, in a finite
universe, die tossing should be replaced by a deterministic algorithm of
sufficient complexity. A spectrum of
different approaches to the problem of time, some of them similar to
the one presented above, can be found in
Ref. \cite{griff}. In a recent paper J. Schneider \cite{schn94} proposes that
a passing  instant is the production of a meaningful symbol,
and must be therefore formalized in a rigorous way as a transition. He
also states that the linear time of physics is the counting of the passing
instants, that time is linked with the production of meaning and is
irreversible per se. We agree only in part, as we strongly believe that
physical events, and the information that is gained due to these events,
are objective and primary with respect to secondary mental or semantic
events.
\subsection{What is classical?}
We consider now the question: {\em what is classical?}\,.
In each practical case, when we want to explain a given phenomenon,
it is clear what constitutes {\em events} for us that we want to
account for. These events are classical, and usually we can safely
extend the classical system $C$ towards $Q$ gaining lot and loosing
little. But here we are asking not a practical question, we are
asking a fundamental question: what is {\em true} $C$? There are
several possibilities here, each one having advantages and disadvantages,
depending on circumstances in which the question is being asked.
If we believe in quantum field theory and if we are ready to take its
lesson, then we must admit that one Hilbert space is not enough, that
there are inequivalent representations of canonical commutation
relations, that there are superselection sectors associated to
different phases. In particular there are inequivalent
infrared representations associated to massless particles
(cf. \cite{roep70} and references therein).
Then
classical events would be, for instance, soft photon creation and
annihilation events. That idea has been suggested by
Stapp \cite{stapp85a,stapp85b}
some ten years ago, and is currently being developed in a rigorous, algebraic
framework by D. Buchholz \cite{buch94a,buch94b}. \\
Another possibility is that not only photons, but also long range
gravitational forces may take part in the transition from potential to
the actual. That hypothesis has been expressed by several authors (see
e.g. contributions of F. K\'arolyh\'azy et al., and R. Penrose in
\cite{penr85}; also L. Diosi \cite{diosi89}).
The two possibilities quoted above are not satisfactory when we
think of a finite universe, evolving step by step, with a finite
number of events. In that case we do not yet know what is gravity
and what is light, as they, together with space, are to emerge only
in the macroscopic limit of an infinite number of events. In such a
case it is natural to look for $C$ in $Q$. We could just {\em define}
event as a nonunitary change of state of $Q$. In other words, we would
take for ${\cal S}_c$ the only available set - the unit ball of the
Hilbert space. This possibility has been already discussed in \cite{ja94a}.
That choice of   ${\cal S}_c$ is also necessary when we want to discuss
the problem
of objectivity of a quantum state. If quantum states are objective (even
if they can be determined only approximately), then the question:
\lq what is the actual state of the system\rq\, is a classical
question - as an attempt to quantize also the position of $\psi$ would
lead to a nonsense. We should perhaps remark here that our picture
of a fixed $Q$ and fixed $C$ that we have discussed in this paper
is oversimplified. When attempting to use the PDP algorithm to create
a finite universe in the spirit of space--time code of D. Finkelstein
(cf. \cite{fink88} and references therein), or bit--string universe of
P. Noyes (cf. Noyes' contribution to \cite{penr85})
we would have to allow for $Q$ and $C$ to grow with the number of events.
Our formalism is flexible enough to adjust to such a change.
\subsection{What is $V$?}
The next question that we have asked is {\em what is $V$?}\, . To
answer this question we must first know the answers to the two previous
questions. In practical situations. when $C$ is specified, then $V$
is chosen so that it provides the best fit to the experimental data.
There are simple rules to construct $V$ and we have discussed
in details several explicit examples in the already quoted references.
On the other hand, when $C$ is related to the infrared representations -
we do not know
the answer yet, but we can see several ways of attacking this problem,
and we hope to return to this case in the future.
When $Q$ is finite and ${\cal S}_c$ is the unit ball in the Hilbert
space - so that we deal with a \lq natural and  universal\rq\, $C$,
then there is also a natural and universal $V$. Indeed, an event is
then simply a pair of state vectors, $|\psi>,|\psi\prime>$, and to
such a pair we can canonically associate the operation $g_{\psi\psi^\prime}=
|\psi ><\psi^\prime|$. That natural  choice defines $V$ up to
a numerical coupling constant. We remark that in this case ${\cal S}_c$
is infinite and continuous, so that the simplified mathematical framework
that we have outlined is insufficient and must be extended.
That this can be easily done was
demonstrated in \cite{blaja93c,ja94a}. In the continuous case the diagonal
of the $V$ matrix is of measure zero and as such - unimportant. But the
conditional expectation ${\cal E}$ of Sect. 2 must be regularized. It
is to be however remarked that what is {\em natural}\/ from
pure mathematical point of view, is usually oversimplified or wrong
when applied to a physical problem. Therefore in any practical problem
the universal $C$ is too big, and the natural $V$ is too simple.
\subsection{Dynamics and Binamics}
Having provided tentative answers to some of the new questions, let us pause
to discuss possible conceptual implications of the EEQT. We notice that
EEQT is a dualistic (and even syncretistic) theory.
In fact, we propose to call the part of time
evolution associated to $V$ by the name of {\em binamics} -- in contrast
to the part associated to $H$, which is called $dynamics$. While
dynamics deals with the laws of exchange of forces, binamics deals with the
laws of exchange of bits (of information). We believe that these two sets of
laws refer to different projections of one reality and neither one
of these projections can be completely reduced to another one. Moreover,
concerning the reality status, we believe that \lq bits\rq\, are as real as
\lq forces\rq\,. That this is indeed the case should be clear if we apply
the famous Lande's  criterion of reality: real is what can kick. We
know that information, when applied in an appropriate way, may cause
changes and may kick - not less than a force.
\subsection{What are {\em we}\, ?}
We have used the term \lq {\em we}\rq\/ too many times to leave it without a
comment. Certainly {\em we}\/ are partly $Q$ and partly $C$
(and partly of something else). But not only we
are subjects and spectators - sometimes we are also actors. In
particular we can {\em gain and utilize information} \cite{gell89}. How
can this happen? How can we control anything? Usually it is assumed that
we can prepare states by manipulating Hamiltonians. But that can not
be exactly true. We are not in power to change coupling constants or
Hamiltonians
that are governing fundamental forces of Nature. And when we say that
we can manipulate Hamiltonians, we really mean that we can manipulate
{\em states}\/
in such a way that the standard fundamental Hamiltonians act on these
special states
{\em as if they were} phenomenological Hamiltonians with classical control
parameters and external fields that we need in order to
explain our laboratory procedures.
So, how can we manipulate states without being able to
manipulate Hamiltonians? We can only guess what could be the answer of
other interpretations of Quantum Theory.
Our answer is: we have some freedom in
manipulating
$C$ and $V$. We can not manipulate dynamics, but binamics is open.
It is
through $V$ and $C$ that we can feedback the processed information
and knowledge - thus our approach seems to  leave comfortable space
for IGUS-es. In other words, although we can exercise  little if any
influence on the continuous, deterministic evolution\footnote{
Probably the influence through the damping
operators $\Lambda_\alpha$ is negligible in normal circumstances},
we may have partial
freedom of intervening, through $C$ and $V$, at bifurcation points, when
die tossing takes place. It may be also remarked that the fact that more
information
can be used than is contained in master equation of standard quantum
theory, may have not only engineering but also biological significance.
In particular,
we provide parameters ($C$ and $V$) that specify event processes that
may be used in biological organization and communication. Thus in EEQT, we
believe, we overcome criticism expressed by B.D. Josephson concerning
universality of quantum mechanics \cite{jos88,jos91}) . The interface
between Quantum Physics and Biology
is certainly also concerned with the fact that a lot of biological
processes (like the emergence of naturally catalytic
molecules or the the evolution of the genetic code) can be
in principle described and understood in terms of physical quantum events of
the kind that we have discussed above.

We believe that are our proposal as outlined in this paper and elaborated
on several examples in the quoted references is indeed the minimal
extension of quantum theory that accounts for events. We believe that,
apart of its practical applications,
it can also serve as a reminder of existence of new ways of looking at old
but important problems.

\end{document}